\documentclass{aa}

\usepackage{newtxtext}
\usepackage{color}
\usepackage{amsmath}
\usepackage{soul}

\DeclareSymbolFont{matha}{OML}{txmi}{m}{it}
\DeclareMathSymbol{\varv}{\mathord}{matha}{118}
\begin{document}

   \title{How the formation of Neptune shapes the Kuiper belt}

   \subtitle{}

   \author{Simona~Pirani\inst{1}
          \and
          Anders~Johansen\inst{1,2}
          \and
          Alexander~J.~Mustill\inst{2}
          }

    \institute{Centre for Star and Planet Formation, Globe Institute, University of Copenhagen, Øster Voldgade 5-7, 1350 Copenhagen, Denmark.\\
   \email{simona.pirani@sund.ku.dk}
   \and
   Lund Observatory, Department of Astronomy and Theoretical Physics, Lund University, Box 43, 22100 Lund, Sweden.\\
   }

   \date{}


  \abstract
  {Inward migration of giant planets is predicted by hydrodynamical simulations during the gas phase of the protoplanetary disc. The phenomenon is also invoked to explain resonant and near-resonant exoplanetary system structures. The early inward migration may also have affected our Solar System and sculpted its different minor planet reservoirs. In this study we explore how the early inward migration of the giant planets shapes the Kuiper Belt. We test different scenarios with only Neptune and Uranus and with all the four giant planets, including also some models with the subsequent outward planetesimal-driven migration of Neptune after the gas dispersal. We find objects populating mean motion resonances even when Neptune and Uranus do not migrate at all or only migrate inwards. When the planets are fixed, planetesimals stick only temporarily to the mean motion resonances, while inwards migration yields a new channel to populate the resonances without invoking convergent migration. In these cases, however, it is hard to populate mean motion resonances that do not cross the planetesimal disc (such as 2:1 and 5:2) and there is a lack of resonant KBOs that cross Neptune's orbit. These Neptune crossers are an unambiguous signature of the outward migration of Neptune. The starting position and the growth rate of Neptune matters for the contamination of the classical Kuiper belt region from neighbouring regions. The eccentricity and inclination space of the hot classicals and the scattered disc region become much more populated when all the giant planets are included. The 5:2 resonance with Neptune becomes increasingly populated with deeper inward migrations of Neptune. The overall inclination distribution, however, is still narrower than from observations, as is generally the case for Kuiper belt population models.}


   \maketitle
%

\section{Introduction}\label{sec:intro}

\begin{figure}
\begin{center}
\includegraphics[width=\hsize]{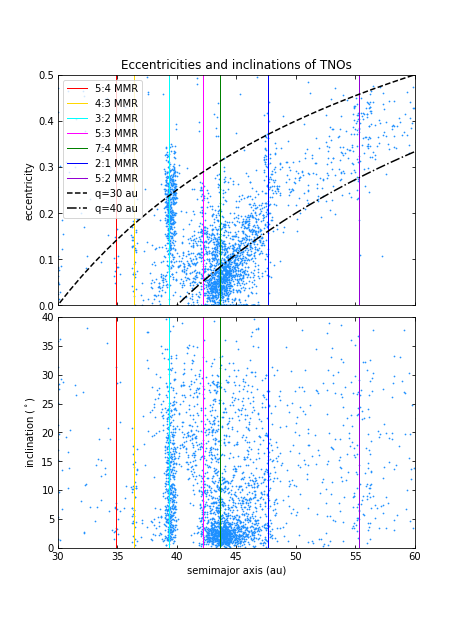}
\caption[]{Eccentricities (upper figure) and inclinations (bottom figure) of TNOs. Data are from the IAU Minor Planet Center. The dashed line and the dash-dotted line represent respectively \textit{q}=30 and \textit{q}=40 au perihelion distances. Vertical solid coloured lines indicate the main mean-motion resonances with Neptune.}
\label{fig:TNOs}
\end{center}
\end{figure}

The Kuiper belt (KB) is a population of small bodies that extends from Neptune's orbit, at about 30 au, to roughly 50 au in the trans-Neptunian region. Its mass is estimated to be of the order of $10^{-2}$ $M_\oplus$ \citep{gladman01,bernstein04,pitjeva18,diruscio20}. Overlapping the outer edge of the Kuiper Belt, there is a second region called the scattered disk, which continues outward to about 1000 au. Figure \ref{fig:TNOs} shows the eccentricities versus semimajor axis (top plot) and the inclinations versus semimajor axis (bottom plot) of the trans-Neptunian objects (TNOs) in the region between 30 au and 60 au. The data are from the IAU Minor Planet Center\footnote{\url{https://minorplanetcenter.net/data}} database. As the figure displays, the region is surprisingly composed of various dynamical classes and this diversity is believed to be linked to the outer Solar System's early dynamical history. The main dynamical classes are \citep{gladman08}:
\begin{enumerate}[(i)]
\item The classical Kuiper belt (CKB). These are objects orbiting with semimajor axis between the 3:2 and the 2:1 external mean motion resonances with Neptune, respectively at about 39 au and 47.5 au. However, there is a lack of low inclination objects between the 3:2 MMR and 42 au, due to the destabilizing effect of the $\nu_8$ secular resonance. The CKB objects are further subdivided into:
\begin{itemize}
\item The cold classicals (CCs), with  $i<5^\circ$ and $e<0.1$. This component does not stretch all the way out to the 2:1 resonance, but depletes quickly after 45 au \citep{kavelaars09}. These objects typically have red optical colours \citep{gulbis06} with $\approx$30$\%$ of the population found in binaries \citep{grundy11}. The cold component comprises $\approx$50$\%$ of the CKB total population \citep{petit11} and it is believed that it has formed in situ, due to the presence of extremely wide binaries that otherwise would have been destroyed by close encounters with Neptune \citep{parker10}.
\item The hot classicals (HCs), with a wider range of inclinations and eccentricities compared to CCs. They represent the other $50\%$ of the total CKB population \citep{petit11}\footnote{It is worth mentioning that work by \citet{fraser14} estimates that the \textit{hot} population is about 30 times more massive than the \textit{cold} one. However, the definitions of \textit{hot} and \textit{cold} classes differ from the CCs and HCs classes. The \textit{cold} objects are defined as having $i<5^\circ$ and semimajor axis in between 38 au $\le$ a $\le$ 48 au, while \textit{hot} objects are from 30 au $\le$ R $\le$ 150 au (a much larger region compared to the HCs one) with $i>5^\circ$.}. Contrary to the CCs, HCs have probably been implanted in their current location, evidenced by the fact that the two populations are very different from each other. Indeed, besides more excited orbits, the HCs show a wider range of photometric colours and a lower binary fraction than CCs \citep{noll08}.
\end{itemize}
\item The resonant KBOs are objects in mean motion resonance with Neptune, such as Plutinos (3:2), Twotinos (2:1), 5:3, 7:4 and 5:2 populations, etc. The resonant objects estimate is about $25\%$ of the total Kuiper belt population \citep{gladman12}. \citet{gulbis06} found that objects in the 3:2 resonance span the full range of KBO colours. The objects in the 4:3, 5:3, 7:4, 2:1 resonances are predominantly red and the 5:2 resonant objects are grey (or neutral) colour. Moreover, the 2:1 resonance inclination distribution is much colder than the 5:2 and 3:2 resonances \citep{volk16}.
\item The scattered disc, that extends from 30 au to roughly 1000 au, interacts with Neptune by close encounters and hence the scattered disc objects (SDOs) have perihelion distances in between 30 au and roughly 40 au. They are mainly grey (or neutral) in colour \citep{sheppard10}.
\item The detached objects, instead, have slightly larger perihelion distances than the scattered disc objects and are no longer affected by close encounters with Neptune.
\end{enumerate}

The first hypothesis about the very peculiar structure of the Kuiper belt was proposed by \citet{malhotra93,malhotra95} who predicted the existence of objects in external mean-motion resonances with Neptune. In the model, they are trapped via adiabatic resonance sweeping during a phase of outward planetesimal-driven migration of the ice giant \citep{fernandez84}. Indeed, resonance capture can occur when the orbital frequency of a planetesimal and/or planet changes gradually and the resonance "sweeps" across the orbit of the planetesimal. Resonance sweeping may result in the capture of the planetesimal in the resonance (if the migration direction is convergent and sufficiently slow) or in exciting the planetesimal to higher eccentricity without capture taking place, when the migration is divergent \citep{henrard83}. Since then, a myriad of works on how the outward migration of Neptune can shape the trans-Neptunian region have been carried out \citep{chiang03,gomes03,hahn05,levison08,nesvorny15a,nesvorny15b,nesvorny16}. The accepted paradigm for HCs is that they formed inside $30$ au and were implanted into the CKB region because of the outward migration of Neptune. CCs are thought to be formed in situ, at $> 40$ au, and did not get much disturbed by the migration of Neptune, explaining the high rate of wide binaries in the population. Recent models seem to prefer a Neptune migration that was slow, long-range and grainy \citep{nesvorny15a,nesvorny15b,nesvorny16}. Nevertheless, the rich structure of the Kuiper belt is not easily reproducible. The main issues can be grouped in three topics: the colour problem, the inclination problem and the resonant population problem.

\textit{The colour problem.} CCs have their own red colour (in literature often called \textit{very red} or \textit{ultra-red}), distinct from the dynamically excited population (HCs, resonant KBOs and SDOs with $i>5^\circ$) that shows a broad range of colours \citep{pike17}. Also, the overall colour distribution of dynamically excited TNOs is found to be bimodal and can be divided into a grey (or neutral) class and a red class. Red dynamically excited objects have lower inclinations ($5^\circ<i<20^\circ$) and grey ones have an inclination range much wider. The different orbital distributions of the grey and red dynamically excited TNOs provide strong evidence that their colours are due to different formation locations in a disc of planetesimals with a compositional gradient \citep{marsset19}. This is potentially in agreement with the planetesimal-driven migration of Neptune because the higher inclinations of the grey (or neutral) dynamically excited TNOs imply that they have experienced more interaction with Neptune than the red ones \citep{gomes03}, suggesting a grey--red--(CC)red composition of the primordial planetesimal disc. On the other hand, \citet{fraser17} reported the detection of a population of neutral-coloured, tenuously bound binaries residing among the CCs. They suggested that these binaries may be contaminants originated at about $\approx$38 au that are shown to be able to survive push-out into the cold classicals during the early phases of Neptune's migration. \citet{schwamb19}, based on optical and NIR colours of 35 TNOs and on the presence of neutral wide binaries in the CCs, hypothesised a protoplanetesimal disc with red--grey--(CC)red composition, with the excited red class closer to the Sun, the grey class (or neutral) in between $\approx$33 and $\approx$40 au and the red cold classicals beyond 40 au, in contrast with \citet{marsset19}. Hence, TNOs colours have a layer of complexity still to disentangle.

\textit{The inclination problem.} A long-standing problem of the different migration models of Neptune is that all of them, more or less, produce a narrower inclination distribution of Plutinos and HCs compared to observations \citep{brown01,gomes03,nesvorny15a}. The inclinations may have been excited prior to Neptune's migration, but no such early excitation process has been identified so far \citep{chiang03}. The inclinations are also not strongly correlated with the timescale of Neptune's migration \citep[see][]{nesvorny15a,volk19}. The lack of cold Plutinos is also very puzzling: classical models reproduce this feature by truncating the planetesimal disc at about 34 au and letting Neptune `jump' during the late instability of the giant planets \citep{levison08}.

\textit{The resonant population problem.} Finally, models with a smooth migration of Neptune predict excessively populated resonances compared to the CKB population that is estimated to contain more mass than Plutinos \citep{gladman12}.  \citet{nesvorny16} showed that a grainy migration of Neptune cannot resolve the overpopulation problem of the resonances compared to the HCs, but produces better results for the ratios between the population in resonances, i.e. the ratios between 3:2, 2:1 and  5:2 \citep[see][Figure 3]{malhotra19}. The 5:2 resonance, in particular, is very hard to populate with current models. It is a third order resonance, located at about 55 au and it is expected to be relatively weak compared to the first order resonances 2:1 and 3:2, but its population is probably comparable with Plutinos \citep{gladman12,volk16}. The concentration of their eccentricity near 0.4 has been explained by \citet{malhotra18} as being related to the resonant width maximum near this value of eccentricity. Nevertheless, how a large number of objects ended up trapped in the 5:2 resonance is still an open question. Indeed, adiabatic resonance capture from an initially cold planetesimal disc is inconsistent with a very populated 5:2 resonance and its inclination distribution \citep{chiang03}. However, the resonance sweeping scenario of \citet{malhotra93,malhotra95} could yield capture efficiencies in the 5:2 resonance approaching those of the 2:1 resonance if Neptune migrated smoothly into a pre-stirred planetesimal disc \citep{chiang03,hahn05}. The alternative is the direct gravitational scattering, but this mechanism leads to large libration amplitudes (exceeding $160^\circ$) for the resonant KBOs, in contrast to observations.

Planet formation models have changed drastically since the discovery of the first trans-Neptunian object in 1993 \citep{jewitt93}, after Pluto and Charon. The main new idea is that the cores of the giant planets grow quickly according to the core accretion model \citep{pollack96} boosted by pebble accretion \citep{johansen10,ormel10,lambrechts12,ida16,johansen17} and migrates inwards due to interactions with the gaseous disc \citep{ward97,lin86}, following growth tracks similar to those shown in \citet{bitsch15b}. This means that the cores of giant planets have to form further away in the outer Solar System with respect to their current location and then migrate inwards while growing. Recently, \citet{pirani19a,pirani19b} showed that the inward migration of the giant planets could have affected also our Solar System. The model, in fact, can explain the Trojan asymmetry very easily: it arises because Jupiter was growing and migrating at the same time during the gaseous phase of the protoplanetary disc. They also showed that, in order to have an asymmetry comparable with the observed one, Jupiter has to migrate at least a few au.
Also from compositional arguments, it has been suggested that Jupiter formed in the outer Solar System. Indeed, in Jupiter's atmosphere, elements like C, N, S, P, Ar, Kr, and Xe are enriched with respect to solar values by approximately a factor of three. If Jupiter accreted around its current location, only some of those elements would have been in the solid phase in the solar nebula, whereas others (Ar and the main carrier of N, N$_2$) would have still been volatile and would have not been accreted onto the core. \citet{oberg19} proposed the idea that Jupiter's core formed beyond $\approx$30 au, where nitrogen and various noble gases were frozen and could be accreted by the core. Then, during envelope accretion and planetesimal bombardment, some of the core mixed in with the envelope, causing the observed enrichment pattern.

If the large-scale migration of the giant planets happened in our Solar System, it could also have shaped the trans-Neptunian region and that is what we are going to explore in this paper. The paper is organised as follows: in section \ref{sec:methods} we describe the \textit{N}-body code, how we built the growth tracks and the different scenarios we tested in our simulations. In section \ref{sec:results} we present our results and analyse the different features of the Kuiper belt obtained with every scenario. Finally, in section \ref{sec:conclusions} we summarise our results and discuss their implications.

\section{Methods}\label{sec:methods}

In this paper, we use simplified version of giant planets’ growth tracks shown in \citet{bitsch15b}, that is we follow the prescriptions presented in \citet{johansen17}. We implemented the growth tracks into an \textit{N}-body code and simulate the evolution of the Solar System over billion of years. We are interested in identifying features of the trans-Neptunuan region that are linked to the early inward migration of the giant planets. Because of this, we also need to mimic the subsequent planetesimal-driven migration of the ice giants and analyse how the shape of the Kuiper belt changes if we include it or not. We will focus on the population of the MMRs with Neptune, the inclinations of the TNOs and the formation location of the planetesimals that end up in the different populations of the trans-Neptunian region.

\subsection{\textit{N}-body code}
In our simulations, we utilised a parallelised version of the \textsc{Mercury} \textit{N}-body code \citep{chambers99} and we selected its hybrid symplectic integrator that is faster than conventional \textit{N}-body algorithms by about one order of magnitude \citep{wisdom91} and particularly suitable for our simulations that involve timescales of the order of billions of years. We used a time step that is 1/20 of the orbital period of the inner planet involved in the simulation \citep{duncan98}. We modified the code so that the giant planets grow and migrate according to the growth tracks generated following the recipes in \citet{johansen17}, as will be explained in subsection \ref{sec:gt}.

In the \textsc{Mercury} \textit{N}-body code, planets are treated as massive bodies, so they perturb and interact with all the other bodies during the integration. The other particles, called small bodies, are perturbed by the massive bodies but cannot affect each other. Since we set them as ``massless'', they also cannot perturb the massive bodies. We will refer to these particles in the text as ``massless particles'' or ``small bodies''. In our simulations we used these massless particles to populate the planetesimal disc in which giant planets grow and migrate.
Our version of the code also includes aerodynamic gas drag effects and tidal gas drag effects to mimic the presence of the gas in the protoplanetary disc as in \citet{pirani19a}. The growing protoplanets are affected by the tidal gas drag and the massless particles are affected by the aerodynamic gas drag from $t=0$ Myr until the gaseous protoplanetary disc photoevaporates at $t=3$ Myr, according to typical disc lifetimes \citep{mamajek09,williams11}. Since small bodies are set to be massless during the integrations, we assign them a radius $r_{\rm{p}}=50$ km  \citep{johansen14} and a density $\rho_{\rm{p}}=1.0$ g/cm$^3$ when computing the effect of the aerodynamic gas drag.

In the \textsc{Mercury} \textit{N}-body, migration and gas drag forces have been added directly as a series of symplectic steps to the integrator that accounts for the evolution of the objects, at the end of each half time step. We switch to asteroidal coordinates, apply the symplectic steps and switch back to cartesian coordinates.

\subsection{Planetesimal disc properties}\label{sec:disc}
\begin{figure}
\begin{center}
\includegraphics[width=\hsize]{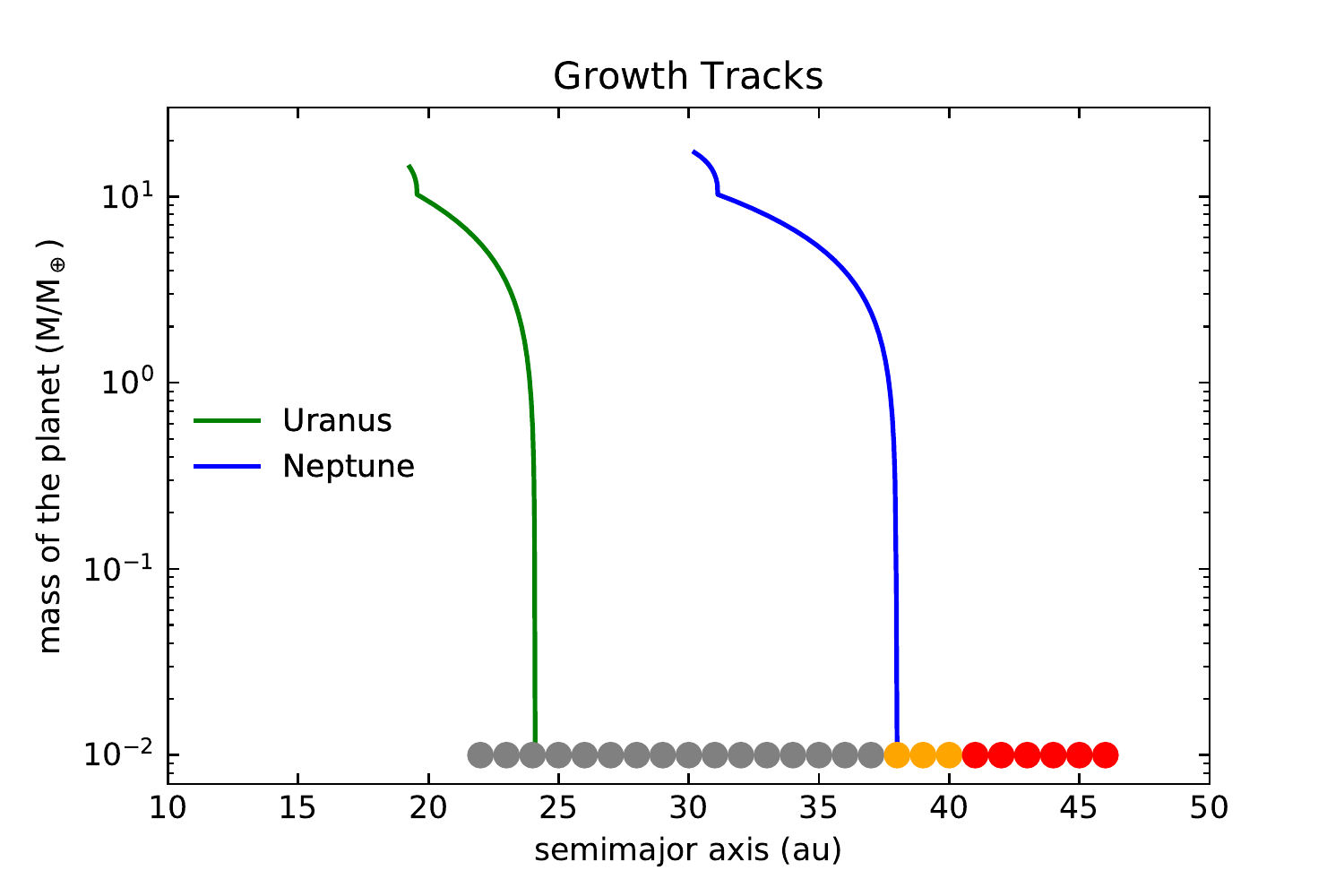}
\caption[]{Growth tracks of Neptune (blue curve) and Uranus (green curve) in our nominal model. In the bottom of the plot the starting colour-coded planetesimal disc.}
\label{fig:growth_tracks}
\end{center}
\end{figure}

We always start with a cold planetesimal disc, with random eccentricities in the interval $[0,0.01]$, random inclinations in the interval $[0^\circ,0.01^\circ],$ and random semimajor axes in each $\Delta a=0.5$ au annular region. We populate annular regions of width $\Delta a=0.5$ au with 1000 massless
particles each. The same amount of particles in each annular
region means a surface density proportional to $r^{-1}$. The outer edge of the planetesimal disc is at 47 au. The inner edge is at 22 au. The small particles are colour-coded in the following way, as shown in Figure \ref{fig:growth_tracks}:
\begin{itemize}
\item Grey for particles initially orbiting inside 38 au;
\item Orange for particles starting in between 38 au and 41 au;
\item Red for particles starting beyond 41 au.
\end{itemize}

We decide to color-code them in this way to mimic the grey--red--(CC)red composition of the primordial planetesimal disc suggested in \citep{marsset19}. In this case, we use the color orange to indicate the red of the dynamically excited population of the KB. The edge between the grey color and the orange color has been chosen so the final SDOs of the different scenarios are predominantly grey. The edge between orange/red colors has been chosen to be 41 au because CCs are supposed to form roughly beyond this limit.

\subsection{No migration}
As a starting point, we decided to explore the structure of the Kuiper Belt region under the influence of the four giant planets in the current configuration.
We used the starting parameters of Jupiter, Saturn, Neptune and Uranus included in the \textsc{Mercury} \textit{N}-body code package. We populated the planetesimal disc from 22 au to 47 au with massless particles: 1000 small bodies every annular region of width $\Delta a = 0.5$ au, for a total of 50000 massless particles. We integrated the simulations for 4.5 Gyr,  analysed which structures form in the trans-Neptunian region and compared them with the ones produced in the other different scenarios. We did not include any gas effect in this particular simulation, since planets start as fully formed and in their current position. They are set in the post-protoplanetary disc phase.

\subsection{How Neptune could shape the Kuiper belt}
As second step, we decided to explore how Neptune, its formation location, and its migration could affect the contamination and excitation of the trans-Neptunian region. In these experiments we include only Neptune and Uranus for two main reasons: (1) the two ice giants are the main influences sculpting the Kuiper Belt region and (2) the strong resonance shifting of the massive gas giants Jupiter and Saturn could easily destabilise the ice giant migration, requiring an accurate fine tuning that would not necessarily add more information for the goal of this section. The full system of the giant planets is added and explored in section \ref{sec:all_planets}.

\subsubsection{Inward migration}\label{sec:gt}
We first tested only inward migration of the ice giants.
In order to generate the ``growth tracks'' for Neptune and Uranus that we implement in our simulations, we used the recipes in \citet{johansen17}. The disc parameters for our model are: $f_{\rm{g}}=0.2$, $f_{\rm{p}}=0.4$, $f_{\rm{pla}}=0.2$, $H/r=0.05$, $H_{\rm{p}}$/$H=0.1$, $\Delta \varv= 30$ m/s and $\textit{St} =0.1$, where $f_{\rm{g}}$, $f_{\rm{p}}$ and $f_{\rm{pla}}$ are parameterisations of the column densities (of the gas, pebbles and planetesimals, respectively) relative to the standard profiles,  $H/r$ is the disc aspect ratio, $H_{\rm{p}}/H$ is the particle midplane layer thickness ratio, $\Delta \varv$ is the sub-Keplerian speed of the gas slowed down by the radial pressure support and $\textit{St}$ is the Stokes number of the pebbles.

The ice giants' growth tracks in the nominal model are shown in Figure \ref{fig:growth_tracks}. The initial mass of the seeds are $10^{-2}$ M$_\oplus$ and they grow until they reach their current masses. The migration starts slightly after 2 Myr and stops when the gas dissipates at 3 Myr.
In the nominal model Neptune's seed starts at 38 au and Uranus' seed starts at 24 au, but we also tested different starting position for the ice giants by modifying the migration rate by a factor larger than one (starting positions of the ice giants further away from the Sun, i.e. 39 au and 40 au) or smaller than one (starting position of the ice giants closer to the Sun, i.e. 37 au, 36 au, 35 au, 33 au). After the disc phase, the ice giant eccentricities and inclinations are artificially increased to reach roughly the current values following similar exponential as Equation \ref{eq:mig} with an e-folding time of $\tau=5$ Kyr.
We analyse the different inward migrations of the ice giant planets and the signature that it would leave on the Kuiper belt.

\subsubsection{Adding the planetesimal-driven migration}

In a second test with only Uranus and Neptune, we included a planetesimal-driven migration \citep{malhotra93,malhotra95} right after the inward migration described in Subsection \ref{sec:gt}.

In order to add the planetesimal-driven migration, we followed the approach adopted in \citet{minton09} and, just after the disc dispersal ($t=3$ Myr), we let Uranus and Neptune migrate following the exponential law
\begin{equation}\label{eq:mig}
a(t)=a_0+\Delta a[1-\rm{exp}(-t/\tau)]
\end{equation}
where $a_0$ is the initial semimajor axis and $\Delta a$ is the final displacement. We also tested a delayed outward migration that starts at 13 Myr, instead of 3 Myr, since it is not exactly established when the outward migration should start and we did want to test is a delay may matter, but we did not find any significant difference in the simulations. We produced growth tracks for Neptune and Uranus where they end their inward migration in an inner orbit compared to the current ones. The final displacement for the subsequent instability was set in order for the ice giants to reach roughly their current semimajor axis, depending on the final position after the inward migration. The parameter $\tau$ is the migration e-folding time. We set $\tau=10$ Myr \citep{nesvorny15a}.

We tested Neptune starting at 39, 37 and 35 au and an inward migration as deep as 21, 24 and 27 au for each different starting position. We obtained the different growth tracks by modifying the migration rate of the planets. Uranus growth tracks are scaled accordingly.

\subsection{Including all the giant planets}\label{sec:all_planets}

We repeated some of the migration simulations including also Jupiter and Saturn. We tested inward migration followed by the planetesimal-driven migration phase.
Jupiter, Saturn, Uranus seeds start at about 18 au, 21 au, 25 au, respectively and reach their current semimajor axis after inward and outward migration. For Neptune's growth tracks, instead, we simulated two different cases: inward migration from (1) 38 to 21 au and from (2) 39 to 18 au, followed by outward migration to its current orbit.


\section{Results}\label{sec:results}

\subsection{Fixed planets}

\begin{figure}[!]
\begin{center}
\includegraphics[width=\hsize]{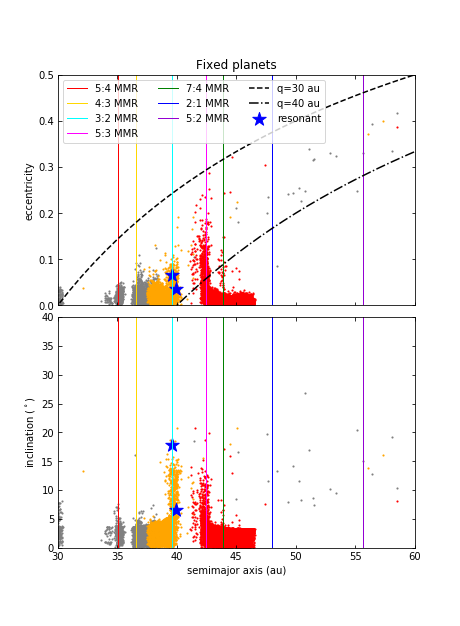}
\caption[]{Osculating eccentricities (top plots) and inclinations (bottom plots) of the trans-Neptunian region after 4.5 Gyr of integration. In this simulation the giant planets do not migrate and they are set with their current orbital parameters. We indicated with blue stars particles that did end up trapped as true resonant in 3:2, 5:3, 2:1 and 5:2 MMRs. The other particles which are in proximity of the MMRs are linked to the phenomenon of \textit{resonance sticking}, that is they are only temporarily trapped in these MMRs and therefore excluded from the true resonant objects group.}
\label{fig:current_nu}
\end{center}
\end{figure}

In this simulation, Jupiter, Saturn, Neptune and Uranus are fixed in their current locations and we let the system evolve for 4.5 Gyr to check the long-term effects on the Kuiper belt. In this case, since the planets start as fully formed and in their current configuration, we do not take into account any gas effects on planets or small bodies.

Results are shown in Figure \ref{fig:current_nu}. As we can see, when the planets do not migrate at all, only sporadic true resonant objects are found. This is not surprising since capture by resonance sweeping requires a convergent migration. With blue stars symbols we display the few particles that are long-term trapped in the 3:2, 5:3, 2:1, 5:2 MMRs with Neptune. We included only particles whose resonant angles\footnote{In order to compute the resonant angle, we used the simplified formula $\phi=p\lambda_{kbo}-q\lambda_N-(p-q)\varpi _{kbo}$, where $\lambda$ and $\varpi$ are the mean longitude and the longitude of perihelion, respectively. $p$ and $q$ are the small integers that relate the orbital periods of the two objects in resonance, $p:q$, with $p>q>0$ for
external resonances.} show a long-term libration, meaning that they undergo bounded oscillations from 3 Myr to the end of the simulations (4.5 Gyr).

\begin{figure}[!]
\begin{center}
\includegraphics[width=\hsize]{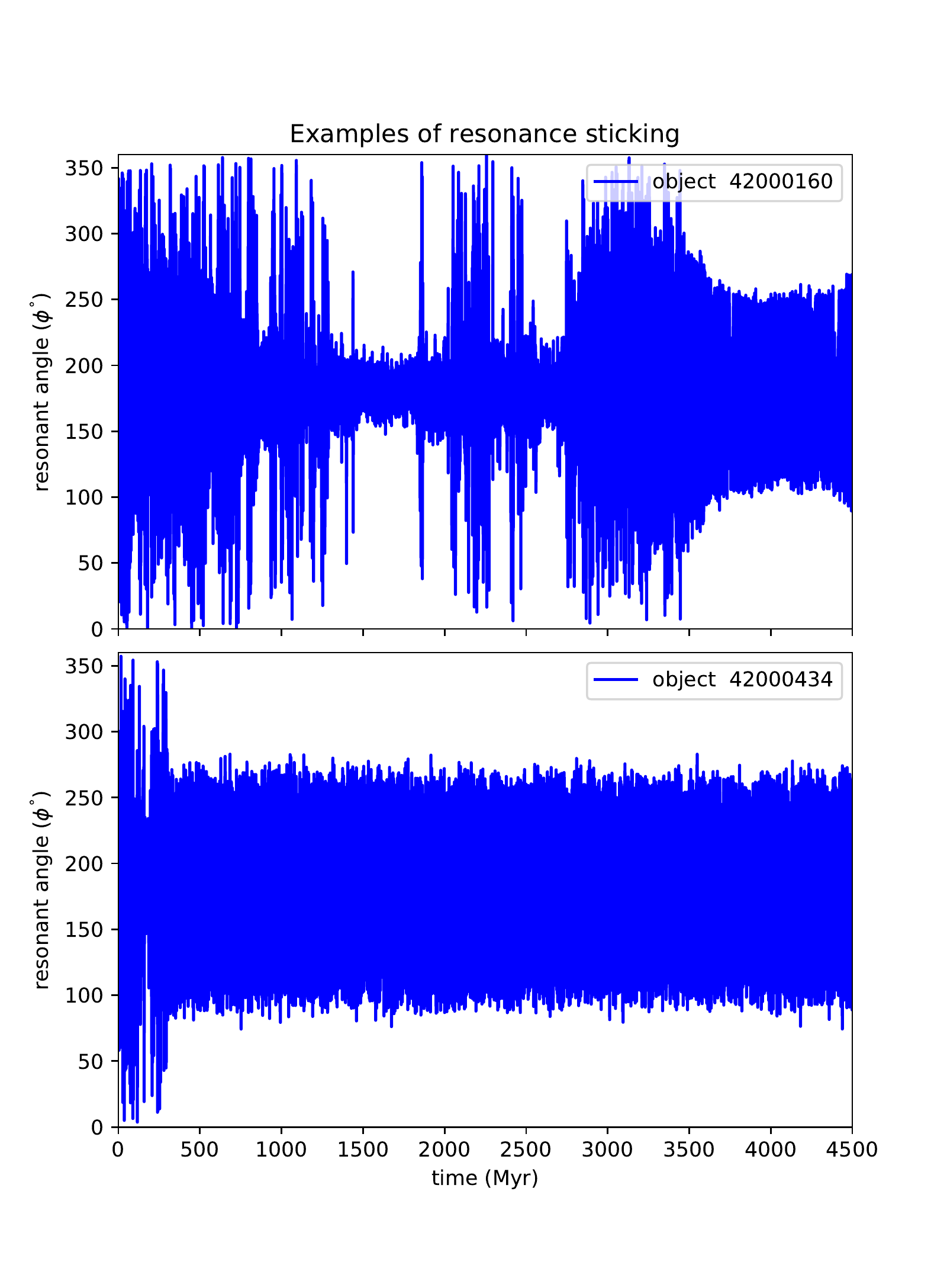}
\caption[]{Example of resonance sticking objects in our simulations. Here $\phi$ is the resonant angle. Objects in a mean motion resonance have values of $\phi$ that librate around a central value. In these examples, however, objects are only temporary trapped, once or multiple times, during the integration time. Some of these objects can show a very long-term sticking to the resonance, as shown in the bottom plot.}
\label{fig:stiking}
\end{center}
\end{figure}
We notice that only 2 of them are really resonant, the other particles in the proximity of a MMR are linked to the phenomenon of \textit{resonance sticking} (see Figure \ref{fig:stiking} for some examples), defined as temporary resonance trapping (single or multiple) during the  object's  dynamical lifetime that can also last for billions of years \citep{duncan97,gladman02,lykawka07}.
When analyzing the resonant arguments of the objects labelled as "true resonant", we notice that these 2 objects start directly in a MMR. As it is expected, no object is trapped in MMR without convergent migration.
The depletion inside 33 au is likely the classical chaotic zone of overlap of first-order MMRs. The plot also shows a gap in between roughly 40-42 au, due to the $\nu_8$ secular resonance with Neptune. Not so many particles ended up as HCs and some resonances, like the 2:1 and the 5:2, do not show a population of resonance sticking objects in contrast to the 3:2 and the 5:3 resonances for the simple fact that there are no planetesimals starting close to those resonance locations. The trans-Neptunian region also presents a scattered disc and few detached objects. We notice that the eccentricities of resonance sticking objects never become large enough so their perihelion distance is closer to the Sun than Neptune, as otherwise common in the observed resonant KB populations. Inclinations are also too low compared to observations. As regards the color-code of the particles, the KB preserves its initial composition gradient with little or no mixing in between the different populations. Only the SDOs present a mix of all the three colours. Finally, we also observe that there is a leftover planetesimal disc with low eccentricities and low inclinations in between roughly 36 au and 39 au. Even if it is not empty nowadays (see Figure \ref{fig:TNOs}), this particular region does not get disturbed enough to get depleted during the simulations and it ends up with too much mass.

\subsection{How Neptune can shape the KB}
\subsubsection{Inward migration}

\begin{figure*}[!]
\begin{center}
\includegraphics[width=\hsize]{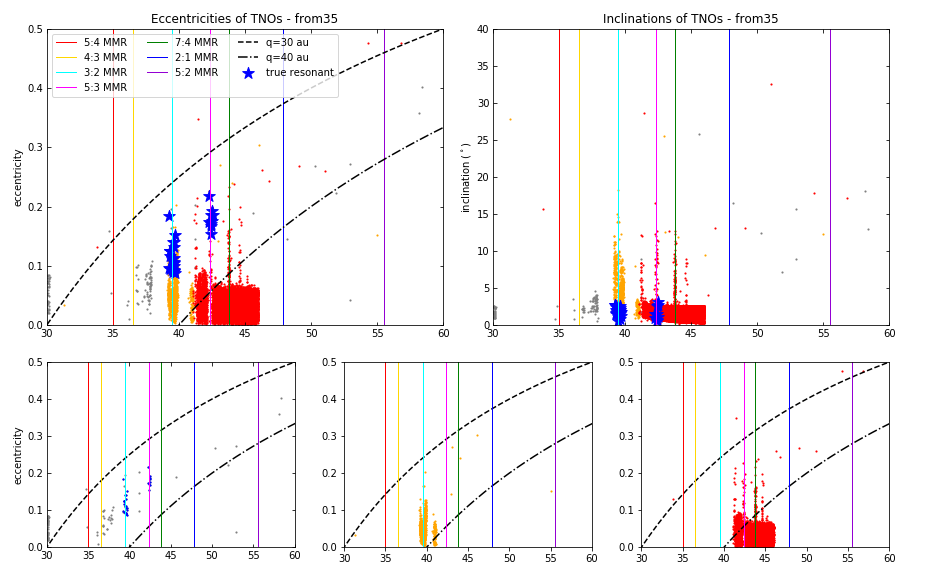}
\includegraphics[width=\hsize]{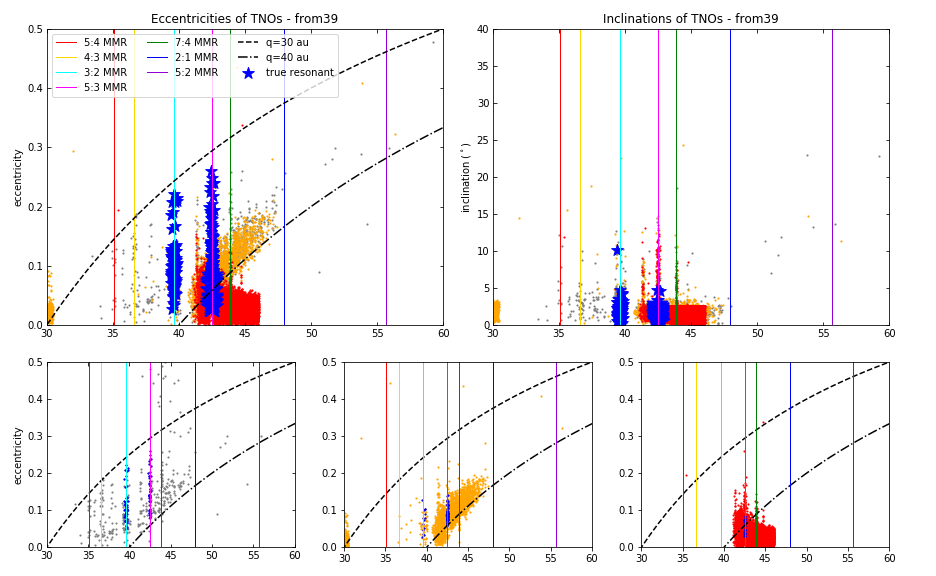}
\caption[]{Osculating eccentricities (left plots) and inclinations (right plots) of the trans-Neptunian region after 4.5 Gyr of integration. Neptune undergoes disc migration from 35 au (top plots) and 39 au (bottom plots) and it reaches its current semimajor axis at about 30 au. Rows containing three plots highlight the separate coloured components. Blue stars/dots indicate true resonant objects. Compared to the "fixed planets" case, the disc in between 34 and 39 au has been depleted by Neptune's inward migration. If Neptune starts very close to the CKB region, some of the orange and grey planetesimals end up forming an low-i/moderate-e clump that overlaps in part with the CCs.}
\label{fig:nsc1}
\end{center}
\end{figure*}

In this set of simulations we consider only the inward migration of Neptune and Uranus during the gas-phase of the protoplanetary disc. After the disc phase, eccentricities and inclinations of the ice giants are artificially increased to reach roughly the current values as explained in Subsection \ref{sec:gt}. We tested different starting positions for Neptune's seed and Uranus is scaled accordingly. Because we consider just the two ice giants, the positions of the secular resonances and their relative gaps may be displaced compared to the actual solar system and their features will be ignored in the analysis. We kept just track of them.


We analysed the results for Neptune starting at 35 au and 39 au and ending its migration in its current position. The other starting positions led to similar results and therefore are not showed here. Snapshots have been taken after 4.5 Gyr of evolution and are shown in Figure \ref{fig:nsc1}. We first studied the objects close to the resonant locations. We expected to find resonance sticking objects and maybe some true resonant objects happened to start the simulation directly in the MMRs as in "fixed planets" case, since the migration is divergent and objects should not be trapped in external MMRs. The results, however, are surprisingly different in this inward migration case. We computed the resonant angles for the objects close to the resonant locations and labelled as true resonant only the ones that have the resonant angle bounded since the dispersal of the disc (3 Myr) till the end of the simulation (4.5 Gyr). We found 31 true resonant objects (22 in the 3:2 MMR and 9 in the 5:3 MMR) in the simulation where Neptune starts at 35 au and 260 true resonant objects (71 in the 3:2 MMR and 189 in the 5:3 MMR) in the simulation where Neptune starts at 39 au. None of them happened to start in a resonance from the beginning.

\begin{figure}[!]
\begin{center}
\includegraphics[width=\hsize]{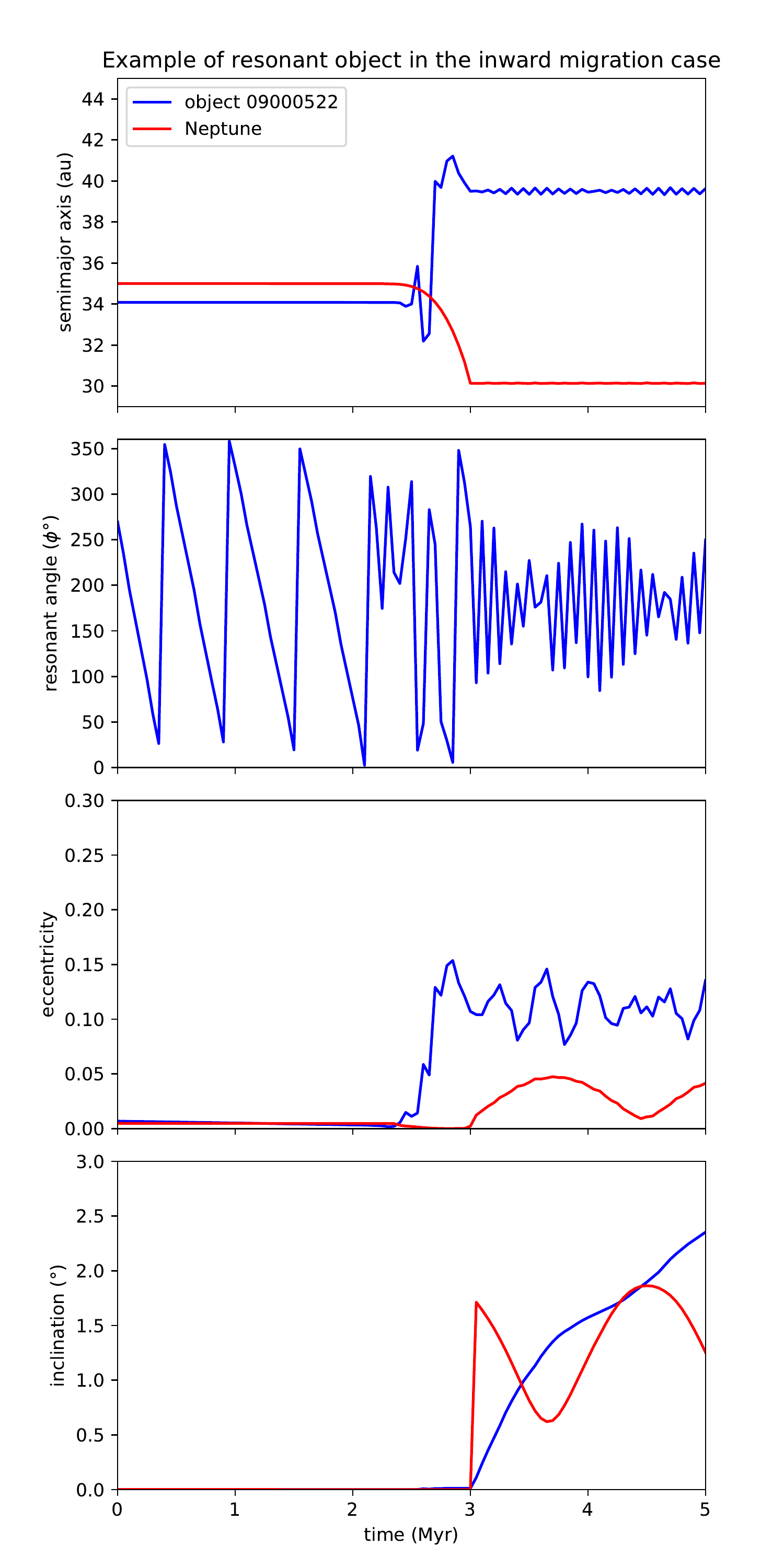}
\caption[]{Semimajor axis, resonant angle, eccentricity and inclination, from top to bottom respectively, of an object in true resonance with Neptune. The red line refers to Neptune's parameters, the blue line refers to the true resonant object's parameters. }
\label{fig:res_inw}
\end{center}
\end{figure}

We analysed semimajor axis, resonant angle, eccentricity and inclination of these objects and Neptune and we show an example of that in Figure \ref{fig:res_inw}. As we can see, these objects are disturbed by the inward migration of Neptune and they are scattered by the planet. What happens next is that they cross or stick to an external MMR when Neptune is increasing its eccentricity. Because of this, the resonance width increases and the object sinks into the resonance. The variety of the libration angle amplitudes of these objects also suggests that this is not a case of resonance sticking, but it is a new trapping mechanism: scattering plus \textit{resonance sinking}. We repeated the same simulation, without artificially increasing the eccentricity and the inclination of Neptune at the end of the disc phase (3 Myr). We found very similar results to the previous case since, at the end of the inward migration, Neptune finds itself close to the 2:1 mutual MMR with Uranus and its eccentricity (that during the disc phase is of the order of $10^{-4}$) increases anyway and starts to oscillates in between roughly 0.001 and 0.01 values in the dance with Uranus.

Analysing the other features present in Figure \ref{fig:nsc1}, we notice that none of the true resonant objects' eccentricities become large enough that their perihelion distance is closer to the Sun than Neptune. Resonances 2:1 and 5:2 are empty since there is less chance that the \textit{resonance sinking} mechanism works where planetesimals are more dispersed. The scattered disc is very underpopulated and some detached objects are also present. The 3:2 resonance is a mix of grey and orange particles. Another very interesting result is that in cases where Neptune starts very close to the CC region, such as 38, 39 and 40 au, grey and orange particles are pushed into the CC region forming population with low inclination and moderate eccentricity that overlaps with the CCs at about 42 au. The disc in between 34 and 39 au that survived in the "fixed planets" case it is way less populated in the inward migration cases because Neptune depletes the region while migrating inwards. There are some other differences in this region depending on where Neptune forms. If Neptune starts at 39 au, the ice giant scatters and excites many more gray objects (that are the ones inside 38 au). This leads to the possibility for them to reach the resonant locations and stick to them or sink in them. The majority of the survivors, in fact, are sticking to the resonances or truly trapped. In case Neptune forms at 35 au, objects are less disturbed by its formation and migration and have less chance to stick to distant resonances or to stick to nearby resonances with higher eccentricity. The gap in between 40 and 42 au is due to the secular resonance with Uranus and not Neptune (as in the current architecture of the solar system), since we did not include all the giant planets in this particular set of simulations. This, however, is not relevant for the scope of this section that is not to reproduce the correct structure of the KBO region, but to investigate how the initial position of Neptune can contaminate the CKBO region.
\subsubsection{Mass left in the planetesimal disc after the large-scale inward migration of Neptune}

\bgroup
\def\arraystretch{1.5}
\begin{table}
\caption{Percentage of the outer planetesimal disc\tablefootmark{(a)} left after the inward migration of Neptune.}             
\label{table:1}      
\centering                          
\begin{tabular}{c c c c c}        
\hline\hline                 

& from35au & from37au & from39au \\

\hline                        
   deep18au & 64.8\%  & 64.9\%  & 63.7\% \\
   deep21au & 48.8\% & 51.4\%    & 52.9\% \\
   deep24au & 33.3\% & 34.2\%    & 37.5\% \\
   deep27au & 33.9\%  & 33.3\%    & 33.1\% \\
\hline                                   
\end{tabular}
\tablefoottext{a}{Planetesimal disc comprised between the position of Neptune prior to the outward migration (18, 21, 24 or 27 au) and 30 au.}

\end{table}
\egroup

In the next section, we analyse a scenario where Neptune migrates first inward during the gas disc phase and then outward mimicking a planetesimal-driven migration because of the presence of the leftover external belt of planetesimals. In order for the planetesimal-driven migration to work, the planetesimal disc outside the orbit of Neptune has to survive the inwards migration of Neptune. Estimates of the mass of the primordial disc of planetesimals outside Neptune's orbit prior the planetesimal-driven migration are of $10 - 30$ M$_\oplus$ \citep{nesvorny12}. In Table \ref{table:1} we summarise our results and show that the disc survives the early inward migration of Neptune (and Uranus) and preserves most of the mass. The surviving disc, in fact, is in the range 64.9\% -- 33.1\% of the primordial one, depending on the different simulations. This means that the inward-outward migration would work for most of the starting mass range proposed for the leftover disc in which the planetesimal-driven migration of Neptune would take place.

Since the particles are massless, for our mass computation, we counted the number of particles left in the disc versus the initial number in the same range of semimajor axis. The inner edge of the disc considered is the deepest semimajor axis reached with the early inward migration (18, 21, 24 and 27 au in the different simulations) when the outer edge is 30 au.
Computations are done at 3 Myr, when the inward migration has just ceased. The following percentages take into account only the planetesimals originally formed inside 30 au, since that is supposed to be the outer edge of the massive leftover planetesimal disc. The contribution of planetesimals formed originally beyond 30 au and ending interior to it, before the outward migration, is considered negligible since beyond $\sim30$ au the number density of the planetesimals per au is at least 2 orders of magnitude smaller than interior to this limit, so that Neptune ceases its outward migration.

\subsubsection{Adding the planetesimal-driven migration}

\begin{figure*}[!]
\begin{center}
\includegraphics[width=\hsize]{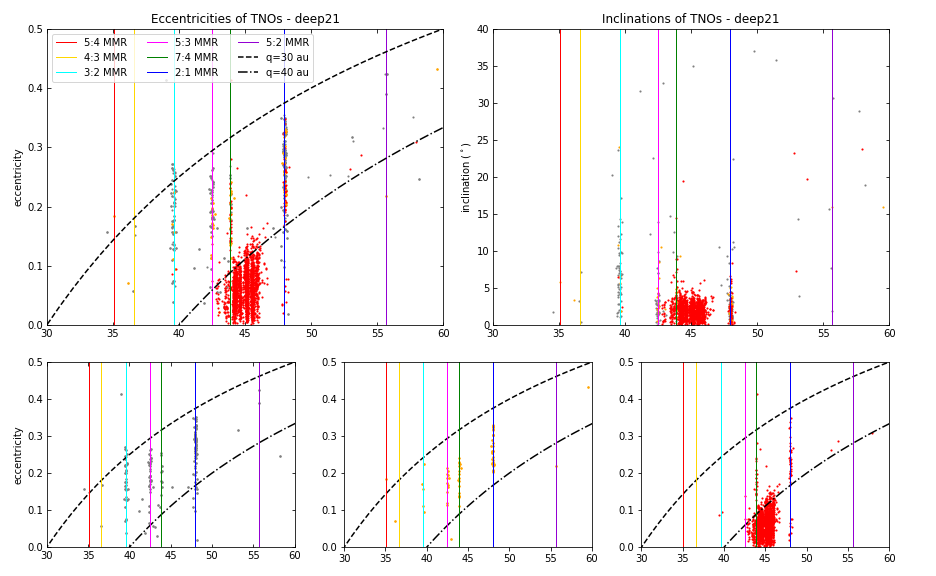}
\includegraphics[width=\hsize]{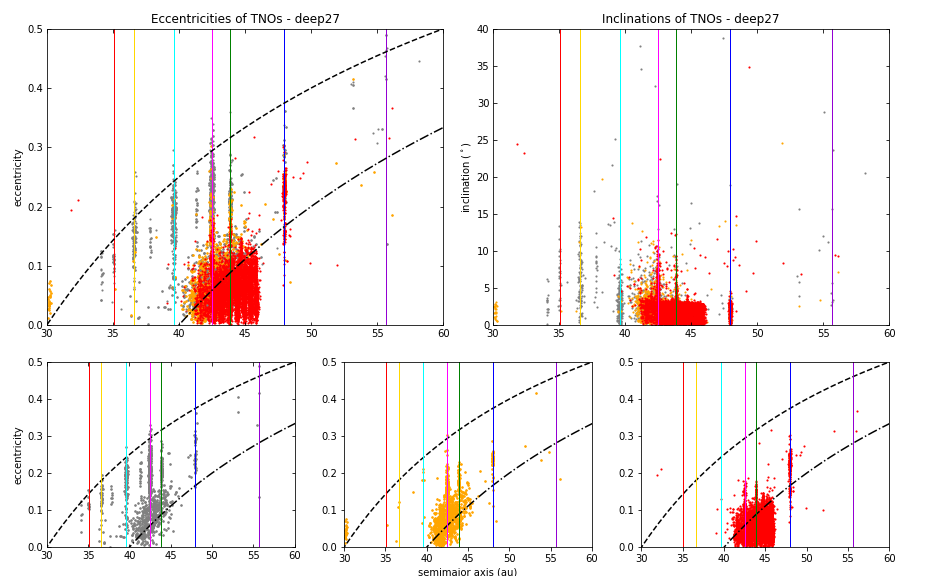}
\caption[]{Osculating eccentricities and inclinations of the trans-Neptunian region after 4.5 Gyr of integration. Neptune undergoes disc migration from 39 au to 21 au (top plot) and 27 au (bottom plot) and then it reaches its current semimajor axis at about 30 au following equation \ref{eq:mig}. Rows containing three plots highlight the separate coloured components. There are some differences, mainly concerning the 3:2, 4:3 and 5:4 resonances. They are less populated if Neptune's migration is as deep as 21 au. }
\label{fig:deep21}
\end{center}
\end{figure*}

In this set of simulations we consider inward disc-migration of Neptune and Uranus followed by an outward migration to their current orbits. In Figure \ref{fig:deep21} we display the eccentricities and the inclinations of objects in the trans-Neptunian region in case Neptune's seed starts at 39 au and undergoes inward migration to a depth of 21 au (top plots) and 27 au (bottom plots) followed by the planetesimal-driven migration to its current orbit. Other cases with different starting positions of the ice giants led to similar results, so we do not show them. The snapshots have been taken after 4.5 Gyr of evolution. As we can see, the MMRs with Neptune are populated, including the 2:1 resonance, in contrast with the previous cases. Nevertheless, resonance 5:2 is almost empty compared to the 3:2 and 2:1 resonances. A scattered disc and few detached objects are present. The HC region is still underpopulated compared to the CCs. The bodies in the 2:1 resonance have lower inclinations compared to 3:2 ones, as is also the case for the observations, but the overall inclination distributions of the MMRs are narrower than observations. The Neptune crossers in the resonances are present in all the combination of starting/final positions we tested, as expected, and this looks like a characteristic that belongs only to the outward migration of Neptune, in the sense that inward migration does not seem to be able to capture resonant objects that are also Neptune crossers. The gap in between 40 and 42 au is present (due to a secular resonance with Uranus, in this particular case with just two planets) and the planetesimal disc interior to 39 au has been depleted by the passage of Neptune in the inward migration phase.

As noticed also in the previous case, where there was only inward migration involved, grey and orange particles overlap with the red ones in the CC region. This is due to Neptune starting at 39 au, that is very close to the CC region. Nevertheless, when Neptune migrates as deep as 21 au, these interlopers are not pushed inside the CC region, since Neptune becomes massive enough to do so when it is already too far away from the CC region. The few objects in the 5:2 resonance are grey, and the SDOs are mostly grey, as in the observed populations. Also the 3:2, 4:3, 5:4 resonances are predominantly grey, in contrast with observations that show that the 3:2 span the full range of KBO colours. The 2:1 resonance is a mix of all the three colours. 

\begin{figure}[!]
\begin{center}
\includegraphics[width=\hsize]{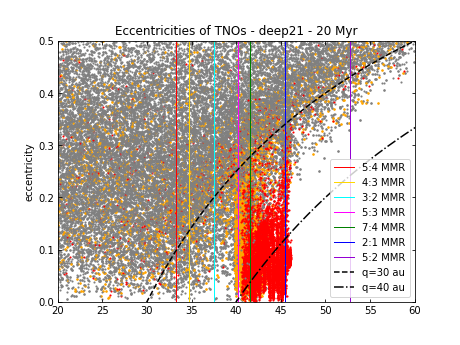}
\includegraphics[width=\hsize]{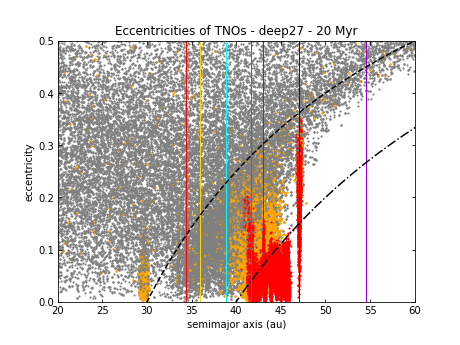}
\caption[]{Osculating eccentricities and inclinations of the trans-Neptunian region at 20 Myr. Neptune undergoes disc migration from 39 au to 21 au and 27 au and at 20 Myr it almost reached its final position at about 30 au following equation \ref{eq:mig}. It can be notice that in the deep21 case, the the degree of excitation of the belt is larger and, for example, the red objects that reach higher eccentricity are scattered when they become Neptune's crossers. In the deep27 case, red objects did not reach enough eccentricity to be scattered around. Resonances (and Neptune's position) are slightly shifted even if snapshots are taken at the same time since the displacement of the outward migration is different in the two cases and so the rate of outward migration.}
\label{fig:deep_20}
\end{center}
\end{figure}

There is a trend of increasing depletion in the KB in simulations deep27, deep24, deep21, deep18 (deep24 and deep18 not displayed in this paper). As shown in Figure \ref{fig:deep_20}, this is due to the fact that the deeper is the inward migration, the more MMRs will cross the CKBO region in the outward migration phase, leading to excitation and depletion (because of the close encounters with Neptune) of the objects that do not have the right \textit{e} and \textit{i} to be trapped via resonance sweeping in resonances that widens only for a narrow parameter space. So, the deeper the inward migration is (or the more long-range the planetesimal-driven migration is), the less populated are the main resonances, since there are less available planetesimals to be captured. In the case where Neptune migrates as deep as 21 au, resonances 5:4 and 4:3 are basically empty compared to when it migrates deep to 27 au.

\subsection{Including all the giant planets}

\begin{figure*}[!]
\begin{center}
\includegraphics[width=\hsize]{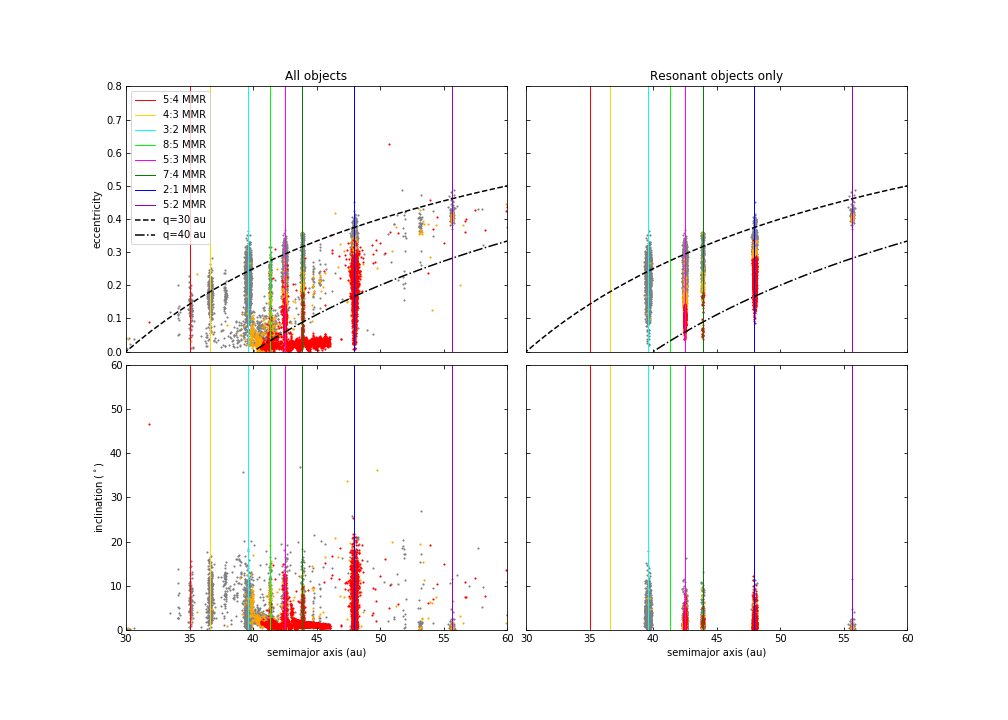}
\caption[]{Osculating eccentricities (top plots) and inclinations (bottom plots) of the trans-Neptunian region after 4.5 Gyr of evolution in the case where Neptune starts from 38 au and migrates inwards to 21 au. All the four giant planets undergo inward disc-migration plus outward migration to their current orbits. In the left plots only true resonant particles in the 3:2, 5:3, 7:4, 2:1 and 5:2 MMRs are displayed. Other resonances have not been checked. We can notice that the 5:2 resonance is populated in this case, even if not as much as the other main MMRs. The plots also show an inclination-color and an eccentricity-color correlation of the particles.}
\label{fig:deep_all}
\end{center}
\end{figure*}

In the last set of simulations, we included all the giant planets: Jupiter, Saturn, Uranus and Neptune and we tested the inward-outward migration of the giant planets. Figure \ref{fig:deep_all} shows snapshots at $t=4.5$ Gyr of the simulations where Neptune starts at 38 au and migrates inwards to 21 au and then outwards to its current location. The outward migration of Neptune, in this particular simulation, starts 10 Myr after the disc dissipates, i.e. at 13 Myr. The results do not differ if we start the outward migration at 3 Myr. Top plots show the eccentricity and bottom plots the inclinations. Plots of the right show only true resonant objects with the main MMRs: 3:2, 5:3, 7:4, 2:1 and 5:2. Other resonances have simply not been checked.

If we analyse Figure \ref{fig:deep_all} we notice that, as expected, resonant KBOs that cross Neptune's orbit are present. The scattered disc is well defined and detached objects are present. Many resonances are populated, way more than the previous cases analysed. The 5:2 resonance is underpopulated and the few objects in the resonance are mainly grey-orange.
\begin{figure}[!]
\begin{center}
\includegraphics[width=\hsize]{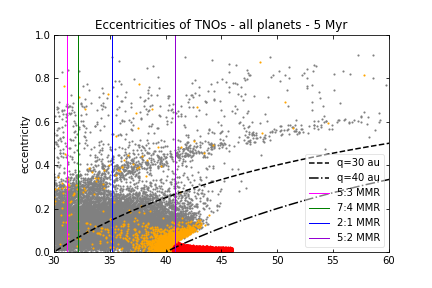}
\includegraphics[width=\hsize]{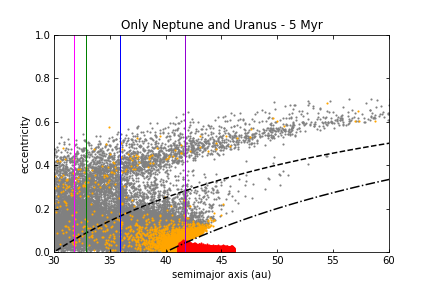}
\caption[]{Comparison between the simulation including all the giant planets and simulation deep21 (from 39 au), where there were only Neptune and Uranus present. The snapshots have been taken at 5 Myr, when the disc phase (and so the inward migration) has already ended and the planets are undergoing the outward migration, as highlighted by the different locations of the MMRs. The simulations diverge during the outward migration due to the enormous scattering efficiency of Jupiter.}
\label{fig:compare}
\end{center}
\end{figure}
The HC region is more populated than the previous cases and the reason lies in the presence of Jupiter scattering objects outwards as shown in Figure \ref{fig:compare} when we compare the snapshot taken at 5 Myr of this simulation with the similar simulation deep21 (from 39 au), where there were only Neptune and Uranus present. There are no significant difference at 3 Myr when the disc dissipates, however the two simulations diverge during the outward migration, due to the enormous scattering efficiency of Jupiter. The overall \textit{i}-distribution is nevertheless still narrower than observations. The 3:2 resonance is mainly grey. The 2:1 resonance is mostly composed of red particles. The resonances in between are a mix of all the three colours. As expected, there is a correlation between eccentricity and colour for the resonant objects. There is also a correlation between colours and inclinations of non-resonant objects, with the grey particles having a broader inclination distribution, followed by the orange ones and the red particles having the narrower inclination distribution. This correlation between inclinations and colours seems to be present also for other populations of the KB. Grey objects are the most excited by the migration of the giant planets and show higher inclinations than orange and red ones. Below roughly $15^\circ$, grey and orange planetesimals seem well mixed.

\begin{figure*}[!]
\begin{center}
\includegraphics[width=\hsize]{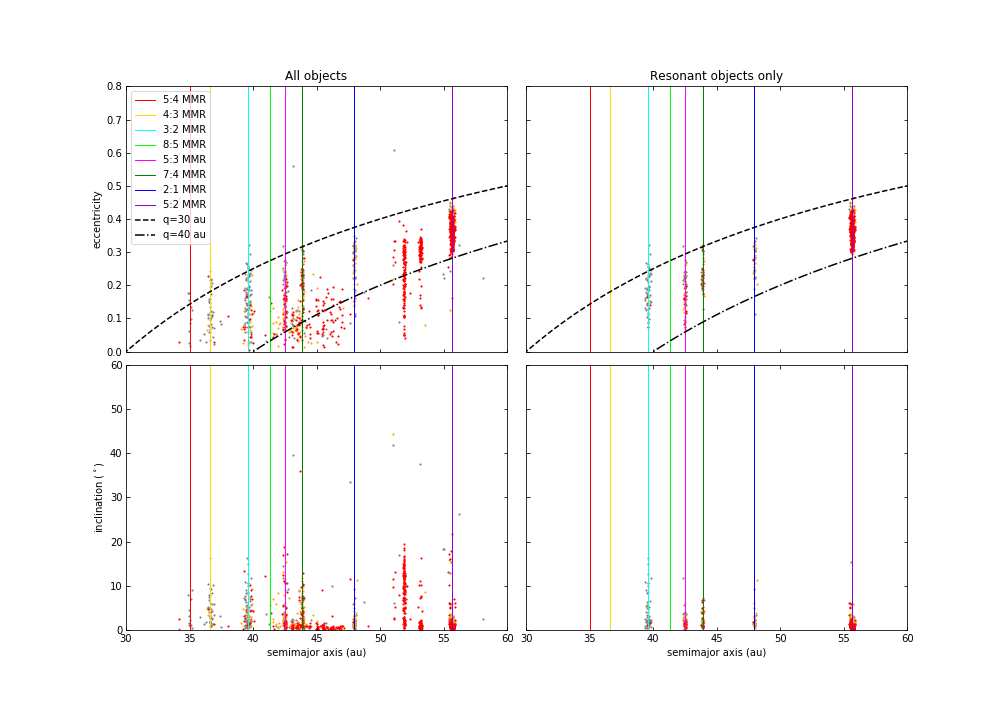}
\caption[]{Osculating eccentricities (top plots) and inclinations (bottom plots) of the trans-Neptunian region after 4.5 Gyr of evolution in the case where Neptune starts from 39 au and migrates inwards to 18 au, then migrates outwards to its current location. All the four giant planets undergo inward plus outward migration to their current orbits. In the right plots only resonant particles in the 3:2, 5:3, 7:4, 2:1 and 5:2 MMRs are displayed. We found that in this case the resonance 5:2 is 6 times more populated than the 3:2 MMR. Moreover, the CKB is depleted compared to the previous cases, but so are the resonant populations.}
\label{fig:deep18}
\end{center}
\end{figure*}

In the case in which Neptune undergoes inward migration from 39 to 18 au and then outward migration to its current location, results are shown in Figure \ref{fig:deep18}.
The plots are really different from the previous case where Neptune migrated only to 21 au. A lot of particles end up in the resonance 5:2, that is 6 times more populated than the 3:2 MMR and mostly composed of red particles. This is because the 5:2 resonance sweeps, before the other resonances, the region roughly in between 33 and 55 au, depleting the region in which the other main MMRs will eventually sweep. It is not possible to efficiently capture bodies in the 5:2 resonance from an unperturbed disc (see \citet{chiang03} or the recent review by \citet{malhotra19}) because of the shape of the resonance. As explained in \citet{chiang03}, the capture probability only increases if the disc is prestirred. When Neptune migrates inwards to 21 au as in the previous scenario, the 5:2 resonance crosses a disc that is almost completely unperturbed, since Neptune forms at 39 au and the stirred region is very narrow. In this simulation, instead, Neptune inward migration is deeper and the 5:2 resonance goes as deep as 33 au crossing the region already stirred by the formation of Neptune and the red objects region (swept and excited by many other resonances), leading to a very efficient capture in the subsequent convergent migration. The CKB is depleted, but so the resonant populations are. Moreover, other weaker MMRs in between 2:1 and 5:2 are found populated. Resonance 3:2 is composed of mainly grey particles as also 2:1, in contrast with the previous experiment. The gap in between 40-42 au now is correctly due to a secular resonance with Neptune. The correlation between eccentricity and colours in the resonant object is not so clear as the previous case.


\section{Discussions and Conclusions}\label{sec:conclusions}

In this paper, we performed simulations of different scenarios that include inwards migration of the giant planets during the gas-disc phase and analysed how they shape the trans-Neptunian region. We started with the giant planets in fixed positions, then we considered inward migration of the ice giants and we also added planetesimal-driven outward migration to their evolution. We concluded with performing inward migration for all the four giant planets plus outward migration. We colour-coded our planetesimal disc in this way: the gray particles are initially placed inside 38 au; the orange ones in between 38 and 41 au, and the red ones in between 41 and 47 au.
Our main conclusions are:
\begin{enumerate}[(i)]
\item Bodies can resonantly stick to the external mean-motion resonances in the KB in the absence of Neptune migration. In the case of inward migration, in addition to resonance sticking, objects can also end up truly trapped in the MMRs. Nevertheless, Neptune crossers in MMRs are a unique signature of the outward planetesimal-driven migration of Neptune, in the sense that only inward migration does not seem to be able to capture resonant objects that are also Neptune crossers.\\

\item Inward migration suggested an additional channel to populate the external MMRs with Neptune, linked to scattered objects and the increase of Neptune's eccentricity that we called \textit{resonance sinking}. This mechanism does not require convergent migration. \\

\item The 5:2 and 2:1 MMRs with Neptune do not end up populated with true resonant objects in the "fixed planets" scenario since capture by resonance sweeping requires a convergent migration. When we considered only the inward migration of Neptune and Uranus and the new resonance trapping mechanism (\textit{resonance sinking}), there is less chance to be trapped since these two resonances do not cross the region initially populated by planetesimals. The scattered objects are then too dispersed to have the chance to end up in these resonances. \\

\item When we include all the 4 giant planets, the inward migration of Jupiter stirs the objects in the whole Solar System, including the trans-Neptunian ones. The result is that we ended up with a significant fraction of HCs, compared to the cases where we included only Neptune and Uranus.\\

\item The formation location of Neptune and its growth rate may matter for the CKB region. Indeed, if Neptune becomes massive enough when it is still very close to the CKB region, it is able to push particles from the neighboring region into it, creating a population of moderate-\textit{e}/moderate-\textit{i} particles that can overlap with CCs. This could be another possible way to push out neutral-coloured wide binaries, if they survive, into the CC region even before the outward migration would start and when the gaseous protoplanetary disc would still in place.\\

\item The population of the resonance 5:2 depends on how deep the inward migration is. In this paper we analysed a Neptune starting its outward migration from 21 au and from 18 au and the population of the 5:2 resonance was scarce in the first case and 6 times more populated than the 3:2 resonance in the latter case. This is because, in the latter case, the 5:2 MMR resonance sweeps a wider region populated with excited planetesimals.\\

\item Grey objects are more excited by the migration of the giant planets and they end up with the highest inclinations among the color-coded ones. Below roughly $15^\circ$ grey and orange planetesimals seem well mixed. This is a confirmation of \citet{gomes03} and in agreement with the grey--red--(CC)red primordial planetesimal disc \citep{marsset19}.\\

\item Based on the simulations, we found the expected correlation between eccentricity and colour of resonant objects, that reflects the primordial composition of the disc.\\

\item \citet{chiang03} found that in order to trap objects in the 5:2 resonance, the trans-Neptunian planetesimals must have been already prestirred. The deep inward migration of the giant planets seems to be the missing mechanism able to stir the trans-Neptunian region prior to the outward planetesimal-driven migration.
\end{enumerate}


\begin{acknowledgements}
We would like to thank the referee for the comments that helped us to improve our work.
SP, AJ and AM are supported by the project grant `IMPACT' from the Knut and Alice Wallenberg Foundation (grant 2014.0017). AJ was further supported by the Knut and Alice Wallenberg Foundation grants 2012.0150 and 2014.0048, the Swedish Research Council (grant 2018-04867) and the European Research Council (ERC Consolidator Grant 724687- PLANETESYS). The computations are performed on resources provided by the Swedish Infrastructure for Computing (SNIC) at the LUNARC-Centre in Lund and partially funded by the Royal Physiographic Society of Lund through a grant.

\end{acknowledgements}

%
   \bibliographystyle{aa} 
   \bibliography{pirani21} 
%


\end{document}